\documentclass[12pt]{iopart}

\usepackage{amssymb}

\usepackage{graphicx}

\newcommand{\lr}[1]{\ensuremath{\left( #1 \right)}}

\begin{document}

\title[1D simulations of detachment]{The role of particle, energy and momentum losses in 1D
  simulations of divertor detachment}

\author{B D Dudson$^1$, J Allen$^1$, T Body$^1$, B Chapman$^{1,2}$,
  C Lau$^1$, L Townley$^1$, D Moulton$^3$, J Harrison$^3$, B Lipschultz$^1$}

\address{$^1$ York Plasma Institute, Department of Physics, University of York, YO10 5DQ, UK}
\address{$^2$ Centre for Fusion, Space and Astrophysics, University of Warwick, Coventry, CV4 7AL, UK}
\address{$^3$ Culham Centre for Fusion Energy, Culham Science Centre, Abingdon, OX14 3DB, UK}

\begin{abstract}
  A new 1D divertor plasma code, SD1D, has been used to examine the
  role of recombination, radiation, and momentum exchange in
  detachment. Neither momentum or power losses by themselves are found
  to be sufficient to produce a reduction in target ion flux in
  detachment (flux rollover); radiative power losses are required to
  a) limit and reduce the ionization source and b) access low-target
  temperature, $T_{target}$, conditions for volumetric momentum
  losses.  Recombination is found to play little role at flux
  rollover, but as $T_{target}$ drops to temperatures around $1$eV, it
  becomes a strong ion sink. In the case where radiative losses are
  dominated by hydrogen, the detachment threshold is identified as a
  minimum gradient of the energy cost per ionisation with respect to
  $T_{target}$. This is also linked to thresholds in $T_{target}$ and
  in the ratio of upstream pressure to power flux.

  A system of determining the detached condition is developed such that the
  divertor solution at a given $T_{target}$ (or lack of one) is determined by
  the simultaneous solution of two equations for target ion current – one
  dependent on power losses and the other on momentum. Depending on the detailed
  momentum and power loss dependence on temperature there are regions of
  $T_{target}$ where there is no solution and the plasma ‘jumps’ from high to
  low $T_{target}$ states. The novel analysis methods developed here provide an
  intuitive way to understand complex detachment phenomena, and can potentially
  be used to predict how changes in the seeding impurity used or recycling
  aspects of the divertor can be utilised to modify the development of
  detachment.
\end{abstract}

\vspace{2pc}

\submitto{\PPCF}
\maketitle


\section{Introduction}
\label{sec:intro}

It has long been recognised that divertor plasma detachment will be
required in future fusion devices such as ITER and DEMO, in order to keep
divertor target heat loads below technological
limits (e.g.~\cite{ipb-divsol-1999,ipb-divsol-2007,efdaroadmap}). Modelling
of divertor detachment in magnetic confinement fusion devices is often done using 2D
models~\cite{rognlien-1999,chankin-2000,schneider-2006}, but simplified
analytic~\cite{stangeby-2000,krasheninnikov2017} and 1D computational
models~\cite{nakazawa-2000,goswami-2001,nakamura2011,togo2013,havlickova} can provide insight into 
the underlying processes, and provide guidance for optimisation of
future devices. Here the SD1D model is presented (section~\ref{sec:SD1D}), which has
been developed using BOUT++~\cite{dudson2014,dudson-2015} to study detachment
dynamics. It is a time-dependent code, which enables the study of the
detachment process, feedback control of detachment, and the
response to plasma transients such as ELMs, in addition to
steady-state solutions. 

Before applying this model to time-dependent problems, we first use SD1D to
understand the roles of particle, power and momentum loss mechanisms involved in
detached steady state solutions. The importance of power loss vs momentum loss
to the detachment process has been debated in the
literature~\cite{stangeby-2000,krasheninnikov2017,stangeby2018} and studied
experimentally~\cite{lipschultz1999,verhaegh}. In addition, simplified models have been
constructed which emphasise power losses to predict the detachment
threshold~\cite{lengyel1981,hutchinson-1994,lipschultz2016,goldston2017} and
sensitivity of the detached region extent/location to external
controls~\cite{lipschultz2016}. 

All simulations are carried out in MAST-Upgrade like geometry, with a parallel
heat flux of $50$MW/m$^2$ at the X-point, $30$m connection length comprising
$10$m above the X-point and $20$m from X-point to target. These are typical of
expected conditions in the first phase of MAST-Upgrade
operation~\cite{havlickova-2014}. The effect of gradients in the total magnetic
field (total flux expansion~\cite{lipschultz2016,moulton-2017}) is included,
with an area expansion factor of 2 (ratio of the total field at the X-point to
that at the target) between X-point and target in all cases shown here.

\section{The SD1D model}
\label{sec:SD1D}

A 1D time dependent fluid model~\cite{stangeby-2000,nakamura2011,togo2013}
is solved for the plasma density $n$,
parallel momentum $nv_{||}$ and static pressure $p = 2enT$, assuming equal isotropic ion and
electron temperatures $T = T_e = T_i$.
\numparts
\begin{eqnarray}
  \frac{\partial n}{\partial t} &=& - \nabla\cdot\left[ \mathbf{b} v_{||} n\right]  + S_n - S \\
  \frac{\partial}{\partial t}\left(\frac{3}{2}p\right) &=& -\nabla\cdot\mathbf{q}_e + v_{||}\partial_{||}p + S_E - E - R \\
  \frac{\partial}{\partial t}\left(m_i nv_{||}\right) &=& -\nabla\cdot\left[m_inv_{||}\mathbf{b} v_{||}\right] - \partial_{||} p  - F\\
  \mathbf{q}_e &=& \frac{5}{2}p\mathbf{b} v_{||} - \kappa_{||}\partial_{||}T_e ,
\end{eqnarray}
\endnumparts
where $\partial_{||} \equiv \mathbf{b}\cdot\nabla$. As discussed in~\cite{ghendrih2011},
the use of isotropic pressure likely overestimates the magnetic mirror
effect in low collisionality regimes. It is also known experimentally
that in general $T_i > T_e$~\cite{kocan-2008,elmore-2013}. 
More sophisticated models retaining both parallel and perpendicular ion pressures
have been developed~\cite{goswami-2001,togo2015}, together with separate ion and
electron temperatures. Those models used a simpler neutral gas model than is
employed here, discussed below. Here we focus mainly on high collisionality
regimes, and leave removing these limitations to future work.

An external source
of power $S_E$ injects energy at a constant rate into a volume above
the X-point, in this case the first $10$m of the domain. The external
source of particles $S_n$ is varied using a
proportional-integral (PI) feedback controller to achieve a specified
upstream plasma density. Coupling to neutrals occurs through particle
sources and sinks (ionisation and recombination) represented by $S$;
energy exchange $E$; radiation $R$ due to hydrogen excitation and
impurity radiation; and friction forces $F$ due to ionisation,
recombination and charge exchange. These hydrogenic rates
are calculated using semi-analytic approximations~\cite{havlickova,leddy2017}.
Unless otherwise stated, in these simulations a 1\% fixed fraction
carbon impurity model is used, based on coronal equilibrium and
calculated using ADAS data~\cite{adas}. In all results shown here where both
are included, hydrogenic radiation exceeds the carbon impurity
radiation.

A similar set of three equations is evolved for the neutral fluid density $n_n$,
parallel momentum $n_nv_{||n}$ and pressure $p_n = en_nT_n$.
The neutrals are not confined by the magnetic field, so transport of
neutrals across the magnetic field can provide a way for neutrals to
migrate upstream. To mimic this process in a 1D model, the effective
parallel velocity is given by the sum of a parallel flow and parallel projection
of a perpendicular diffusion:
\begin{equation}
v_{n} = v_{||n} - \left(\frac{B_\phi}{B_\theta}\right)^2\frac{\partial_{||}p_n}{\nu} ,
\label{eq:neutral_velocity}
\end{equation}
where the collision frequency $\nu$ includes charge exchange,
ionisation, and neutral-neutral collisions. The factor
$\left(\frac{B_\phi}{B_\theta}\right)^2$ is set to $10$ in all
simulation results shown here.

One of the assumptions made in going to a one-dimensional model concerns the
handling of momentum losses. In particular, fast charge-exchanged
neutrals can leave the thin SOL, transferring their momentum directly
to the walls of the device without interacting with the rest of the
neutral gas. In most of the simulations shown here charged
exchanged neutrals do not escape; momentum is conserved, so that the total
pressure (plasma + neutrals) is constant. 

\section{Detachment development}
\label{sec:detachment}

The definition of detachment varies amongst publications. In this paper our
focus is on the particle flux to the divertor target, $\Gamma$, which goes
through at least two stages during detachment in the context of the simple
equation for sheath ion flux:
\begin{equation}
  \Gamma \propto n_{target}\sqrt{T_{target}} \propto p_{target}/\sqrt{T_{target}} .
  \label{eq:targetflux}
\end{equation}
During the attached phase $\Gamma$ rises $\propto T_{target}^{-1/2}$ as the
target pressure $p_{target}$ is constant. During the first stage of detachment
the rise of $\Gamma$ slows through a combination of ionization source loss and
$p_{target}$ drop. We define the second stage, and the focus of most of this
study, to be the point in any scan (e.g. upstream density, impurity seeding, SOL
power) where $\Gamma$ ‘rolls over’; in other words the rate of change of $\Gamma$ becomes
negative with respect to what is being varied. In this paper there are only
density scans, although the carbon fraction in some cases is held constant
leading to the carbon density rising.

\begin{figure}[h]
  \centering
  \includegraphics[width=0.6\textwidth]{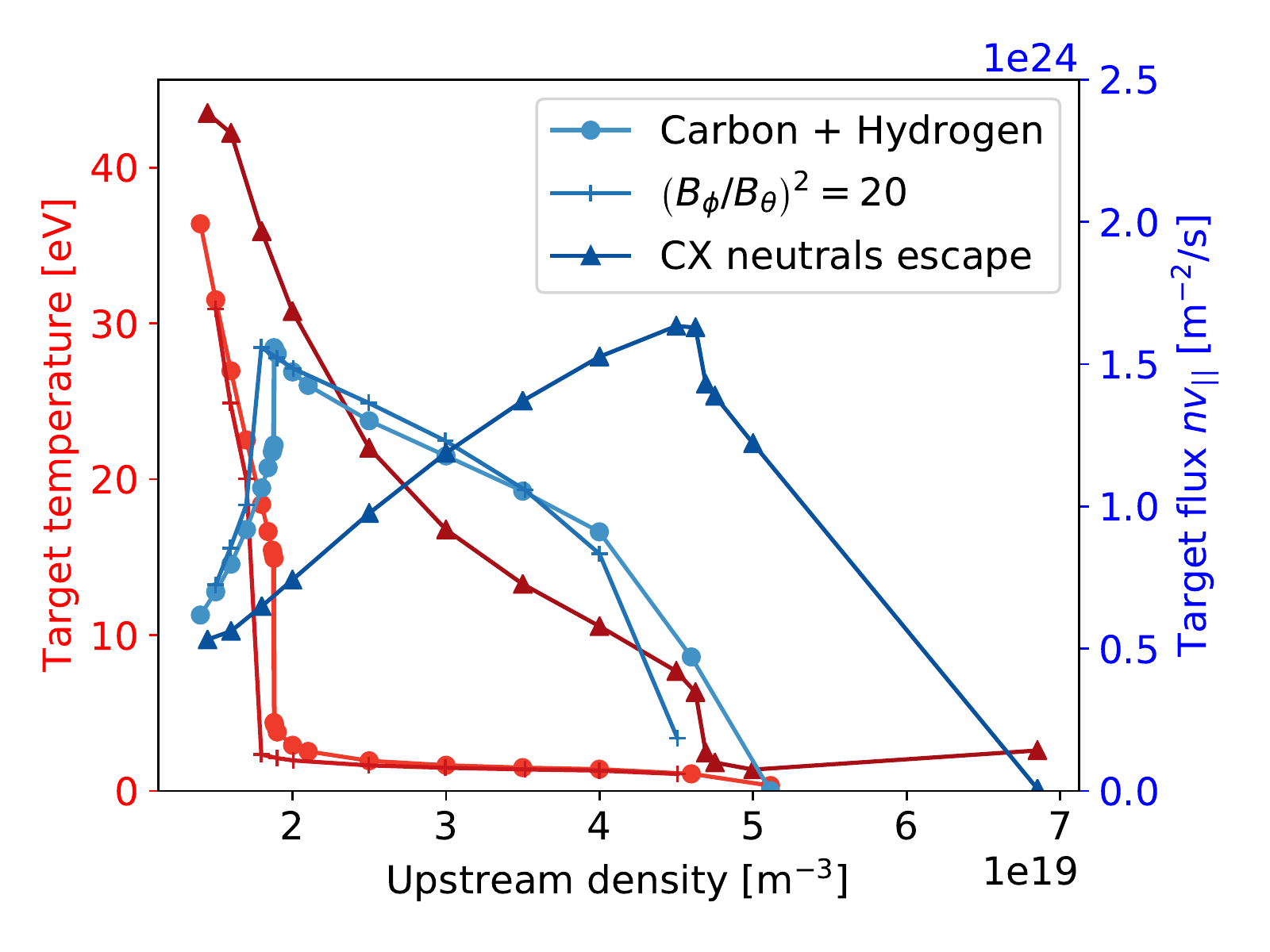}
  \caption{Upstream density scan for a reference case ``Carbon +
    Hydrogen'' with hydrogen excitation radiation and 1\% carbon
    impurity. For comparison are shown a case with doubled cross-field neutral
    diffusion (equation~\ref{eq:neutral_velocity}), and a case in
    which charge exchanged neutrals escape the plasma and return with
    no net momentum (labelled ``CX neutrals escape'').}
  \label{fig:area_2_assumptions}
\end{figure}

The effects of a reference upstream density scan on target ion flux $\Gamma$ are
shown in figure~\ref{fig:area_2_assumptions}, and this scan will serve as a
benchmark for other cases shown in this paper. We find that target ion flux
rollover for the case labelled ``Carbon + Hydrogen'' occurs at an upstream
density of around $1.89\times 10^{19}$m$^{-3}$ for fixed input power flux of
$50$MW/m$^2$. The ``Carbon + Hydrogen'' cases include radiative power losses
from both hydrogen excitation radiation and 1\% carbon impurity.
Also shown in figure~\ref{fig:area_2_assumptions} are results for a case where
the cross-field neutral diffusion is doubled
(equation~\ref{eq:neutral_velocity}, ``$\lr{B_\phi/B_\theta}^2 = 20$''), and a
case in which charge exchanged neutrals remove momentum from the simulation
instead of transferring it to the neutral fluid (``CX neutrals
escape''). Variation in the cross-field diffusion (due to changes in field-line
pitch in equation~\ref{eq:neutral_velocity}) has a modest effect on the results,
but modification to the total (plasma + neutral) momentum loss has a large
effect. This effect of neutrals escaping the SOL is examined further in
section~\ref{sec:momentum}, where it is shown to be important in determining the
total momentum loss fraction $f_{mom}$ (figure~\ref{fig:pratio}). This neutral
loss process will depend on the device and configuration, so an improved model
for neutral momentum loss is likely required for quantitative agreement with
experiment.

In order to understand the
behaviour of this model, and the underlying physical mechanisms,
we here examine the plasma-neutral processes occuring during
detachment. Of particular interest is the role of particle and energy
loss (section~\ref{sec:part_energy}), and 
the role of momentum loss (section~\ref{sec:momentum}) on the flux
rollover at detachment.

SD1D has been benchmarked against the modified two-point
model~\cite{kotov-2009,moulton-2017}, with good agreement for upstream
densities below the flux rollover. As the momentum
and power loss fractions ($f_{mom}$ and $f_{pow}$)
increase, the result becomes sensitive to
small errors in these quantities, so the agreement deteriorates:
relative error is $3\times 10^{-5}$ at $n_{up}=1.8\times 10^{19}$m$^{-3}$
($f_{mom}=0.22$, $f_{pow}=0.25$); 
$2$\% after flux rollover at $n_{up}=1.9\times 10^{19}$m$^{-3}$ ($f_{mom}=0.73$,
$f_{pow}=0.78$); $10$\% at $n_{up}=3\times 10^{19}$m$^{-3}$ ($f_{mom}=0.87$,
$f_{pow}=0.92$); $80$\% error at $n_{up}=4\times 10^{19}$m$^{-3}$ ($f_{mom}=0.91$,
$f_{pow}=0.98$). Some initial comparisons have also been made to the
SOLPS-ITER code~\cite{schneider-2006,wiesen-2015}, with agreement of
the order of $20$\%, but this is the
subject of ongoing work, in particular to understand the
validity of the fluid neutral model used here. Grid convergence tests have been
performed for these simulations, and many of the operators used in SD1D have been
verified using the Method of Manufactured Solutions
(MMS)~\cite{dudson-2015} though not yet the full model.
Simulation results shown here used 800 grid cells, with convergence
tests done at half and double resolution. Grid cells are packed close to the
target so that the resolution parallel to the magnetic field is
$3.8$mm at the target and $7$cm upstream.

\section{The role of particle and energy loss}
\label{sec:part_energy}

The parallel heat flux to the divertor target $q_{||}$ consists of
plasma thermal energy, and the ionisation energy $E_{iz}$ released
by surface recombination. Both of these heat fluxes are proportional
to the particle flux $\Gamma$. Reducing target particle flux is
therefore considered here as the important outcome of detachment for
extending high power divertor lifetime. 

Volume recombination of plasma ions at low temperatures (of the order
of 1eV) provides a mechanism by which plasma ion flux to the
target can be reduced, converting plasma flux to neutral flux before
the plasma reaches the target. To examine the importance of
recombination to these 1D results we have performed the same density
scan but with recombination turned off. 
\begin{figure}[h]
  \centering
  \includegraphics[width=0.6\textwidth]{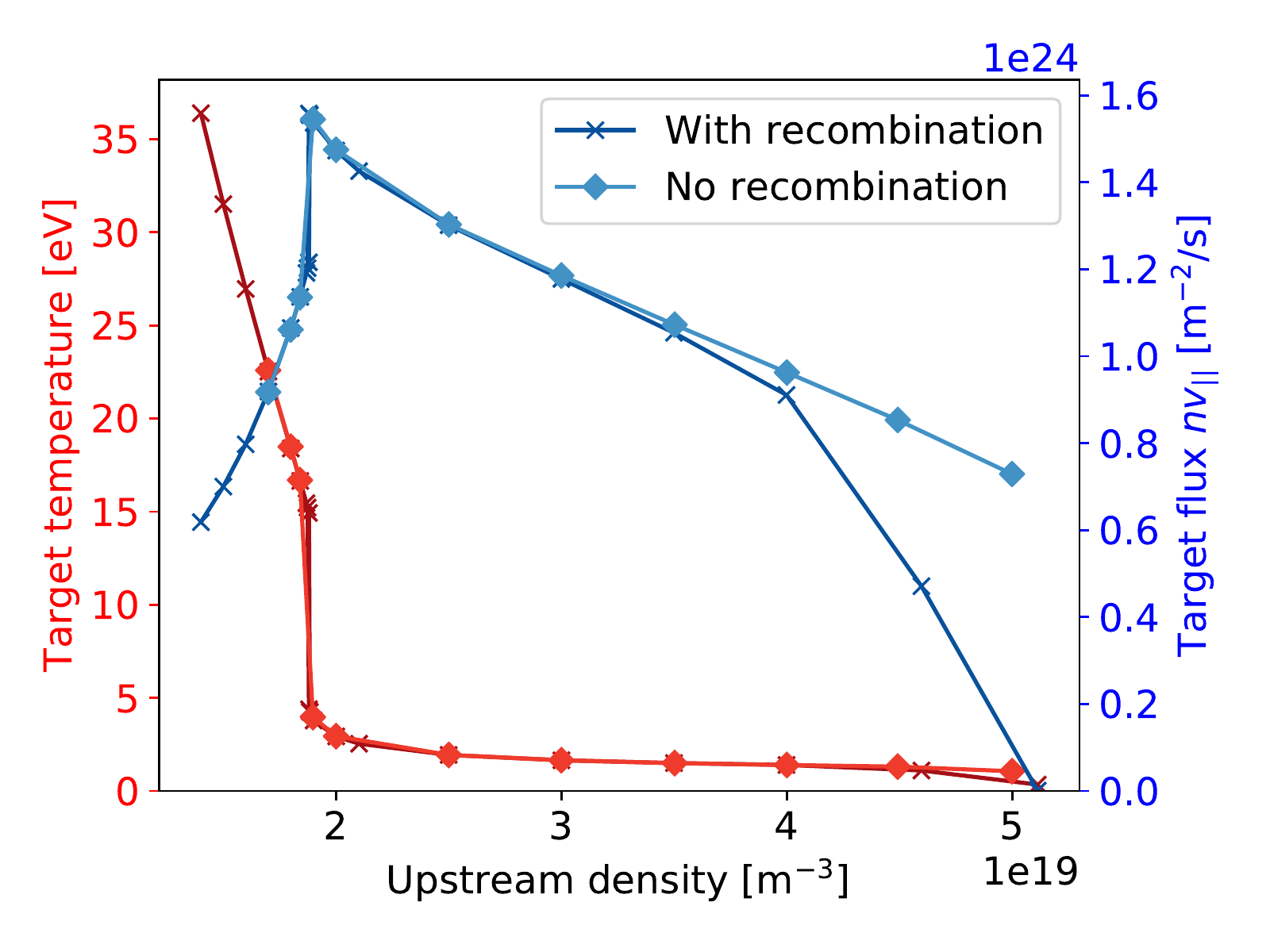}
  \caption{Comparison of density scans with recombination removed.
    The target electron temperature is shown in red, and target particle flux in
    blue. Hydrogen excitation radiation and a 1\% carbon impurity are included.}
  \label{fig:norc}
\end{figure}
Results are shown in figure~\ref{fig:norc}, and indicate that
recombination plays little role at the rollover in flux at an
upstream density of $n_{up} \sim 2\times 10^{19}$m$^{-3}$, but
becomes important at higher upstream densities of around $4\times 10^{19}$m$^{-3}$.
This makes sense because target $T_e \simeq 1.2$eV at this higher upstream density, whereas
at target flux rollover target $T_e \simeq 3.2$eV. This is in contrast
to some previous results~\cite{pschenov2017} in which flux rollover
occurred at lower temperatures, but in agreement with
recent experiments~\cite{verhaegh2017, verhaegh} and
simulations~\cite{fil2018} of TCV. There are several ways in which
target flux can be reduced, recombination being one of them, so
discrepancies may be seen between studies in different regimes.

From particle conservation the target flux is given by the sum of the flux from
the SOL into the divertor and the ionisation flux, less the ions lost to
recombination. Other mechanisms such as cross-field transport may also
contribute, but are not included in this 1D model, and will typically average to
zero over the divertor volume.  In these simulations the recycling coefficient
is set to $f_{recycle} = 0.99$ so the upstream flux is small, and in the absence
of significant recombination the drop in target flux must be due to a drop in
ionisation.  The input power is fixed, so this drop in ionisation is due to
either a reduction in power available for ionisation (due to increased impurity
radiation losses), or an increase in the effective energy cost per ionisation
$E_{iz}$, or both. This effective ionisation cost is calculated as the ratio of
the total power lost to hydrogen ionisation and excitation, divided by the rate
of ionisations.  $E_{iz}$ increases as $T_e$ falls, due to excitation radiation,
in this simulation from $\sim 31$eV per ionisation at
$n_{up}=1.4\times 10^{19}$m$^{-3}$ to $\sim 80$eV per ionisation at rollover
(averaged over the domain; shown in more detail in figure~\ref{fig:eiz_te}).

One way to quantify this power limitation picture of detachment is in
terms of the fraction of power available which goes into hydrogen
excitation and ionisation, with the remainder being transmitted
through the sheath~\cite{verhaegh}. In these simulations the radiation
and ionisation regions overlap, complicating the
analysis. Nevertheless, at flux rollover the fraction of the power
into the divertor going to ionisation increases from $\sim 40$\% to
$\sim 75$\% during the strong drop in $T_{target}$, supporting the
association between power limitation and flux
rollover~\cite{lipschultz1999,krasheninnikov-1999,krasheninnikov2017,pschenov2017,verhaegh}.

\begin{figure}[h]
  \centering
  \includegraphics[width=0.6\textwidth]{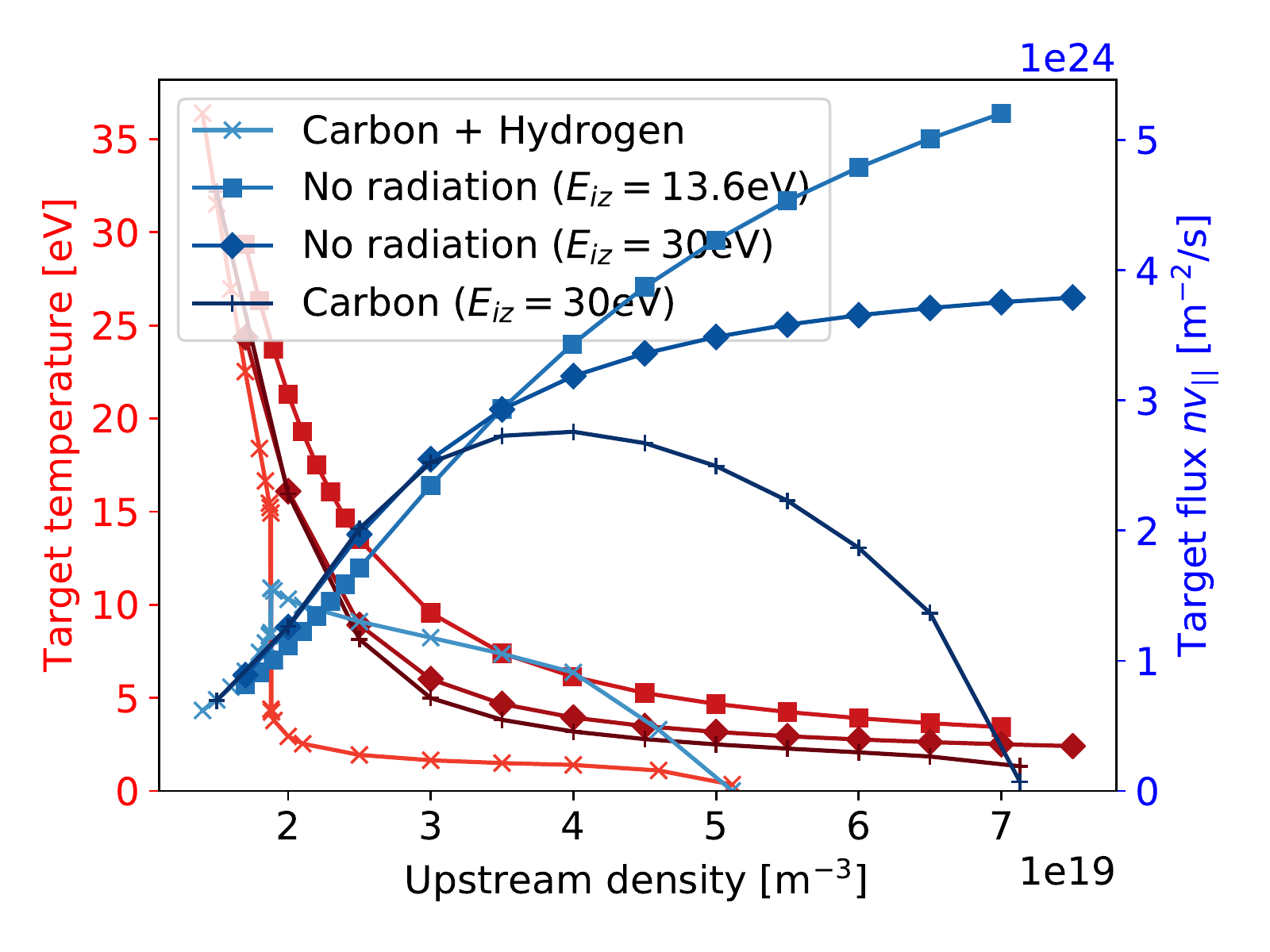}
  \caption{Comparison of density scans with different radiative power
    loss mechanisms, both hydrogenic and impurity: ``Carbon +
    Hydrogen'' is the original case with an effective ionisation cost
    which varies with $T$ ; ``No radiation'' cases have no carbon
    radiation and a fixed ionisation cost; ``Carbon'' has impurity
    radiation and a fixed ionisation cost. The target electron
    temperature is shown in red, and target particle flux in blue.}
  \label{fig:norad}
\end{figure}
To test the importance of power dissipation,
figure~\ref{fig:norad} data labelled ``No radiation ($E_{iz}=13.6$eV)'' shows
the effect of removing all 
power radiation mechanisms, so that the energy cost per ionisation is
only the ionisation potential $E_{iz} = 13.6$eV, and there is no impurity
radiation. In this case rollover during detachment does not occur in the range of
upstream density studied, though the target
temperature falls to $3.3$eV at an upstream density of
$7.5\times 10^{19}$m$^{-3}$, comparable to the target temperature at
flux rollover in the case with radiation included, at an upstream
density of $1.89\times 10^{19}$m$^{-3}$. The fraction of available
power going to ionisation only reaches $38$\% in this case, which
though significant is less than the original case discussed above.

A similar result is found when a fixed ionisation cost of $E_{iz}=30$eV is used,
to account for some hydrogen excitation radiation, also shown in
figure~\ref{fig:norad}.  In this case the fraction of available power going to
ionisation reaches $59$\%. Assuming all of the divertor input power, $P_{in}$, is
used for ionisation, and taking into account the $3.5$eV Franck-Condon energy
with which dissociated neutral atoms are recycled, the upstream input power
would lead to a target flux of
$P_{in}/\left(E_{iz} -3.5\mathrm{eV}\right) = 5.9\times 10^{24}$m$^{-2}$s$^{-1}$
(in the limit of zero target temperature). The highest fluxes reached in the
$E_{iz}=30$eV scan, $\sim 4\times 10^{24}$m$^{-2}$s$^{-1}$, imply that around
$68$\% of the input power is going to ionisation. Though the rate of change of
target flux with upstream density is reducing, no rollover
($d\Gamma/dn_{up} < 0$) in the target flux is seen in this case.  A scan with
$E_{iz} = 60$eV still does not produce a flux rollover.

This lack of flux rollover in cases with fixed ionisation cost can be understood
by considering the power balance and temperature dependence of power losses. In
the absence of significant recombination the target particle flux is given by
the ionisation rate. The input power into the simulation, $P_{in}$, therefore
goes either to impurity radiation ($P_{imp}$), ionisation ($P_{iz}$), or to
target heat flux ($P_{target}$) as given in equation~\ref{eq:powerbalance}
\begin{equation}
  P_{in} = P_{imp} + \underbrace{P_{iz} + P_{target}}_{P_{recl}} = P_{imp} +  \left(E_{iz} + \gamma T_{target}\right)\Gamma^{pow} ,
  \label{eq:powerbalance}
\end{equation}
where $\Gamma$ is the target particle flux, labelled ``pow'' here since this
expression is derived from power balance. The sheath heat
transmission coefficient $\gamma$ is set to $6$ in these simulations, and
$T_{target}$ the temperature at the target (assumed to be the same as at the
sheath entrance). $P_{recl}$ is the power available for ionisation, some of
which is used for ionisation $P_{iz} = E_{iz}\Gamma^{pow}$, and the
remainder going to the target in the form of kinetic energy $P_{target} = \gamma
T_{target}\Gamma^{pow}$. $E_{iz}$ is the energy cost of each ionisation,
consisting of the ionisation potential (13.6eV) and the radiation due to
excitations preceding ionisation. From equation~\ref{eq:powerbalance} we can
write down the power balance constraint for the target ion flux
\begin{equation}
  \Gamma^{pow} = \lr{P_{in} - P_{imp}} / \lr{E_{iz} + \gamma T_{target}} .
  \label{eq:gamma_pow}
\end{equation}
If $E_{iz}$ and $P_{in}$ are fixed, and impurity radiation is negligible, then
with $\Gamma$ cannot be reduced without increasing $T_{target}$, clearly not the
desired outcome. The detached solution with a reduction of both target
temperature and particle flux is therefore not available. This is discussed in
more detail in section~\ref{sec:momentum}.

In this picture there are three ways in which particle flux $\Gamma$ can be
reduced: recombination; an increasing
$E_{iz}$ as $T_{target}$ drops; or impurity radiation $P_{imp}$, which reduces
power available for ionisation. In the original ``Carbon + Hydrogen'' case shown in
figure~\ref{fig:norad} the dominant
mechanism is an increase in $E_{iz}$ (hydrogen radiation) as
$T_{target}$ drops. In the
case labelled ``Carbon $E_{iz}=30$eV'' with a $1$\%
fixed fraction carbon impurity and $30$eV fixed ionisation cost,
particle flux rollover is also achieved, with the fraction of
available power going to ionisation ($P_{iz}/P_{recl}$) around $52$\% at rollover
($n_{up}=3.5\times 10^{19}$m$^{-3}$). High power fractions going to
ionisation appears to be associated with flux rollover, consistent
with the power limitation picture, but is not by itself sufficient to produce
flux rollover. To understand this further we next examine the role of
momentum losses and the description of target ion flux as in
equation~\ref{eq:gamma_pow} but based on momentum ($\Gamma^{mom}$).

\section{The role of momentum loss}
\label{sec:momentum}

As briefly mentioned in section~\ref{sec:detachment}, one of the robust
experimental characteristics of detachment is loss of target plasma
pressure, implying a loss of momentum to neutrals or other sinks.
Simple arguments indicate that the target particle flux varies as
in equation~\ref{eq:targetflux}, reproduced here as
equation~\ref{eq:momentum2pt} and labelled $\Gamma^{mom}$ to indicate that the
origin of this relationship is the momentum balance:
\begin{equation}
  \Gamma^{mom} \propto n_{target}\sqrt{T_{target}} \propto p_{target}/\sqrt{T_{target}} ,
  \label{eq:momentum2pt}
\end{equation}
where the pressure $p_{target}$ and temperature $T_{target}$ are evaluated at
the divertor target.  In the absence of momentum losses, a sonic boundary
condition implies that the static target plasma pressure is proportional to the
upstream pressure $p_{up} = 2p_{target}$. As mentioned, a fall in the target
pressure, faster than the drop in $\sqrt{T_{target}}$, is therefore necessary to
reduce target particle flux, overcoming the increase in flux associated with a
drop in target temperature. We note here that the drop in $p_{target}$ can be
due either to volumetric momentum losses or a drop in the upstream pressure, or
some combination of both~\cite{verhaegh}. For a steady-state solution to exist,
the plasma must self-organise into a configuration which satisfies both this
momentum constraint as well as particle flux and power balance constraints
discussed in section~\ref{sec:part_energy}.

\begin{figure}[h]
  \centering
  \includegraphics[width=0.6\textwidth]{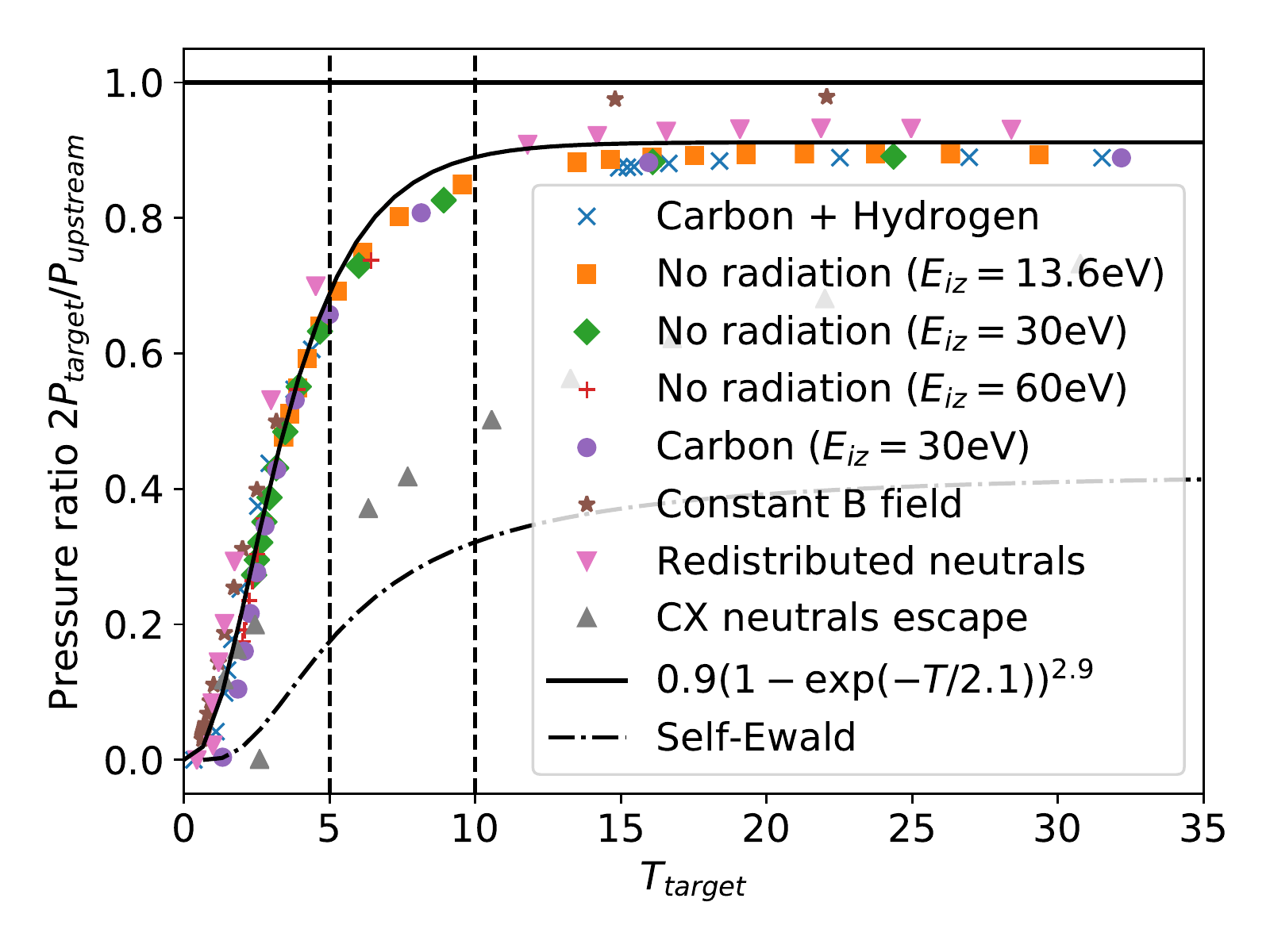}
  \caption{Ratio of target to upstream pressure, as a function of target
    temperature $T_{target}$. The almost universality of the curve indicates
    that this measure of pressure loss is mainly a function of $T_{target}$, in
    agreement with results from experiment and
    modelling~\cite{stangeby2018}. All scans have a total flux expansion factor
    of $2$ from upstream to target except ``Constant B field'' which has no flux
    expansion; all have neutrals recycling at the target except ``Redistributed
    neutrals'' in which 50\% of recycled neutrals are distributed evenly along
    the divertor; all have charge exchange (CX) reactions which conserve
    momentum between plasma and neutral fluids, except ``CX neutrals escape'' in
    which fast neutrals are assumed to leave the plasma, lose their momentum,
    and return at a random point along the divertor.}
  \label{fig:pratio}
\end{figure}
A measure of the pressure loss from X-point to target is the ratio of target
pressure $p_{target}$ to upstream pressure $p_{up}$, shown in
figure~\ref{fig:pratio}. As has been shown
previously~\cite{lipschultz2007,pperez2017,stangeby2018}, this pressure ratio is
found to be mainly a function of target temperature, and here is shown to be
independent of the power dissipation model (Carbon, Hydrogen radiation, or fixed
$E_{iz}$). In addition, a case with no flux expansion (Constant B field), and a
case where 50\% of recycled neutrals are distributed evenly along the divertor
leg (Redistributed neutrals) follow the same trend.  The one case which follows
a different trend in figure~\ref{fig:pratio} is labelled ``CX neutrals escape'',
in which charge-exchanged neutrals escape the SOL, lose their momentum, and
return at a random location along the divertor leg. This results in a loss of
total (plasma + neutral) momentum, modifying $f_{mom}$. As discussed in
section~\ref{sec:SD1D}, the treatment of neutral momentum is likely to be
important in quantatively matching experiment.

As in~\cite{stangeby-2000,stangeby2018}, the pressure ratio and
hence the fraction of total momentum lost, $f_{mom}$, fits a function of the
form in equation~\ref{eq:momlossfit} and shown in figure~\ref{fig:pratio}:
\begin{equation}
  2p_{target}/p_{up} = 1 - f_{mom} = 0.9\left(1 - \exp\left(-T_{target}/2.1\right)\right)^{2.9} \quad .
  \label{eq:momlossfit}
\end{equation}
The fit coefficients found here are comparable to those from experimental
data e.g. C-MOD~\cite{lipschultz2007}
$1-f_{mom}=1.2\left(1-\exp\left(-T_{target}/2.3\right)\right)^{4}$
and SOLPS modelling of AUG~\cite{stangeby2018} H-mode
$1-f_{mom}=0.8\left(1-\exp\left(-T_{target}/2\right)\right)^{1.2}$.
Possible sources of discrepancy include radial momentum transport, and
variations in how plasma momentum is transferred to neutrals and the
walls of the device discussed further below.
As discussed elsewhere~\cite{pperez2017,stangeby2018}, this strong
dependence on target temperature is consistent with a simple model in which 
neutrals ionise in a narrow region close to the target, so that the
neutral density and electron temperature can be considered
homogenous over the interaction region. In this Self-Ewald
model~\cite{self-1966}, given in equation~\ref{eq:self-ewald},
$f_{mom}$ is only a function of ionisation and charge exchange rate
coefficients, which to first approximation depend only on the
temperature:
\numparts
\begin{eqnarray}
  1 - f_{mom} &=& \left(\frac{\alpha}{\alpha + 1}\right)^{\left(\alpha + 1\right)/2} \label{eq:self-ewald} \\
  \alpha &=& \left<\sigma v\right>_{iz} / \left( \left<\sigma v\right>_{iz} + \left<\sigma v\right>_{cx}\right) ,
\end{eqnarray}
\endnumparts
where $\left<\sigma v\right>_{iz}$ and $\left<\sigma v\right>_{cx}$ are
the ionisation and charge-exchange rate coefficients. This Self-Ewald
solution overestimates the momentum loss, as shown in figure~\ref{fig:pratio},
since it assumes that all charge exchange events remove momentum from the system, in addition to occuring in an isothermal environment. 

The drop in pressure from upstream to target shown in figure~\ref{fig:pratio} is
commonly associated with detachment, but is not sufficient to lead to rollover:
Here all scans with conserved momentum follow the same pressure ratio curve, but
the scans with fixed energy cost per ionisation do not have a roll over in
target flux, as shown in figure~\ref{fig:norad}.

\begin{figure}[h]
  \centering
  \includegraphics[width=0.6\textwidth]{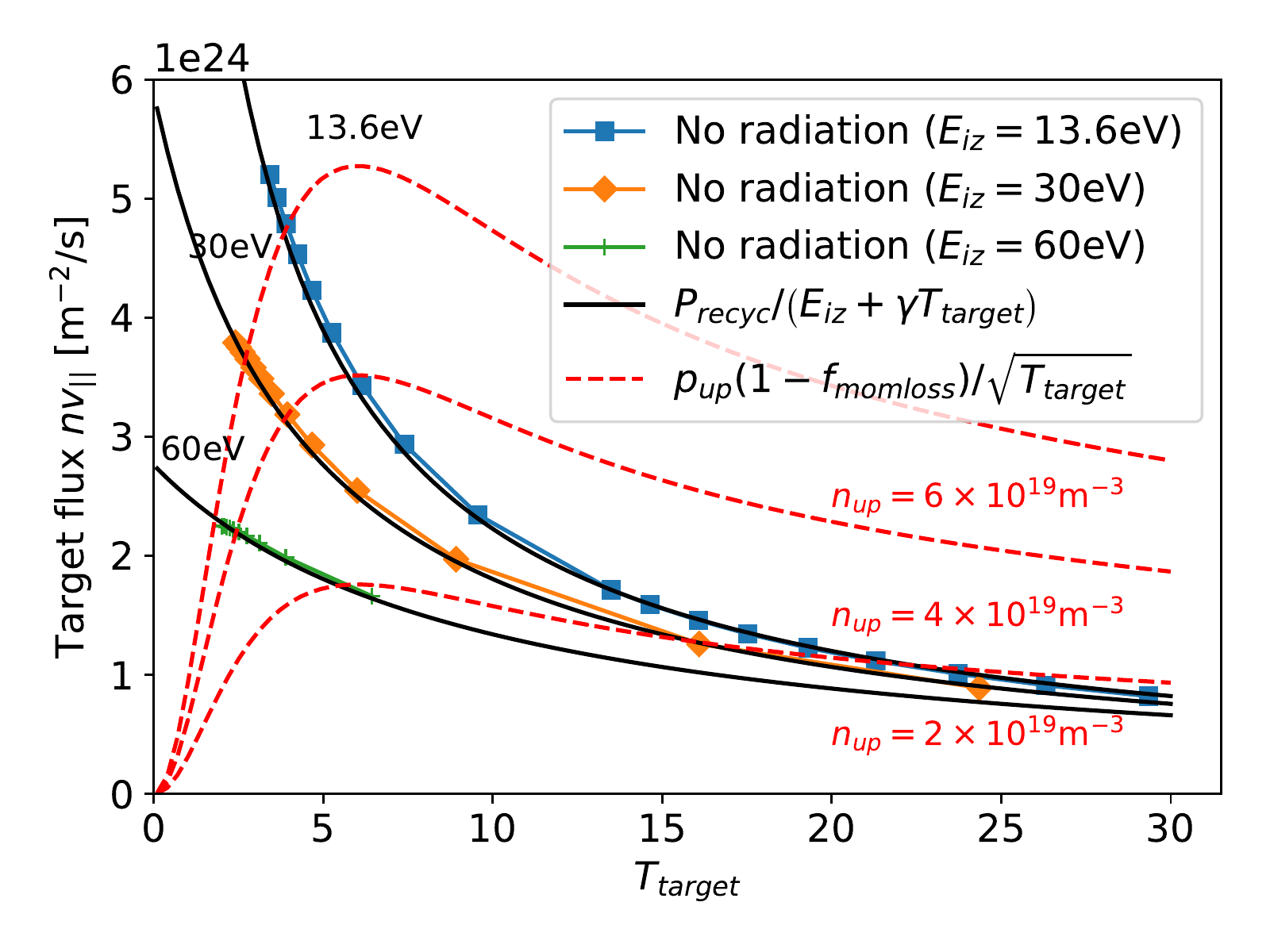}
  \caption{Target particle flux as a function of target temperature
    for three upstream density scans with fixed energy cost per
    ionisation $E_{iz}$ and no impurity radiation, taking into account
    the $3.5eV$ with which neutral atoms are recycled. Black solid curves show
    $\Gamma^{pow}$ calculated using equation~\ref{eq:powerbalance} with
    fixed $P_{in}$ for each value of $E_{iz}$. Red dashed curves
    show $\Gamma^{mom}$ calculated using equation~\ref{eq:gammamom},
    with the fit for $f_{mom}$ from figure~\ref{fig:pratio} and
    equation~\ref{eq:momlossfit}. The intersection of these curves is
    the consistent solution for a given upstream density. Since the
    curves for $\Gamma^{pow}$ are monotonic, no flux rollover is seen
    in these scans.}
  \label{fig:teflux}
\end{figure}

To further understand the lack of rollover in target flux in cases with fixed
energy cost per ionisation $E_{iz}$, figure~\ref{fig:teflux} shows the target
flux against target temperature for three scans with no impurity
radiation and different fixed values for $E_{iz}$.
As discussed in section~\ref{sec:part_energy}, power balance dictates
that the target flux and target temperature 
follows equation~\ref{eq:powerbalance}, shown as solid black lines
for each of the three values of $E_{iz}$. In steady state this
solution must also be consistent with momentum balance
(equation~\ref{eq:momentum2pt}), shown as dashed red curves for three
different upstream densities. The
intersection of the solid black (power) and dashed red (momentum)
curves is therefore the consistent steady-state solution. 

The expression for flux from momentum balance
(equation~\ref{eq:momentum2pt}) can be written as the upstream
pressure multiplied by a function $F\left(T_{target}\right)$ which
only depends on target temperature (equation~\ref{eq:gammamom}):
\begin{equation}
\Gamma^{mom} = p_{up}\underbrace{\frac{\left(1-f_{mom}\right)}{\sqrt{8m_iT_{target}}}}_{F\left(T_{target}\right)} .
\label{eq:gammamom}
\end{equation}
In all SD1D cases discussed here the variation in upstream temperature
is small, $57-61$eV, so $p_{up}\propto n_{up}$. The dashed red curves
in figure~\ref{fig:teflux} therefore scale up and down in proportion
to the upstream pressure ($\sim$ density), whilst retaining a constant
shape $F\left(T_{target}\right)$. The black curves scale in proportion
to the input power, whilst also retaining a constant shape. As these inputs are changed,
the intersection of the two curves moves. Since for large $T_{target}$
the flux from momentum balance (equation~\ref{eq:momentum2pt}) goes
like $\Gamma^{mom} \sim 1/\sqrt{T}$, whilst the flux from power
balance (equation~\ref{eq:powerbalance}) goes like $\Gamma^{pow} \sim
1/T$, there is always a solution for these scans. It can be seen in
figure~\ref{fig:teflux} that as the upstream density is increased the
red curve ($\Gamma^{mom}$) moves upwards; this causes the intersection point to move to
the left,
corresponding to a lowering of target temperature and an increase in
target flux, which asymptotically approaches the limit
$\Gamma_{max}=P_{in}/E_{iz}$. Note that recombination could produce a
rollover in these scans, by removing flux before it reaches the target, but is not
sufficient in the cases studied here.

Having understood why the cases with fixed $E_{iz}$ do not have a
rollover in target flux, we now turn to the more experimentally
relevant cases which do have a flux rollover. From
figure~\ref{fig:teflux} it can be seen that in order to have a
rollover in target flux, the black curve derived from power balance
(equation~\ref{eq:powerbalance}), must change slope so that
$\frac{d\Gamma^{pow}}{dT} > 0$. This is in addition to the requirement
for momentum loss which determines the shape of the red curve.

Assuming a fixed input (upstream) power, $P_{in}$, the power to the
recycling region is given by $P_{recl} = P_{in} -
P_{imp}$ (equation~\ref{eq:powerbalance}), where $P_{imp}$ is the impurity radiation~\cite{verhaegh}. The gradient of
the flux with respect to target temperature is given in equation~\ref{eq:gammapowgradient}.
\begin{equation}
  \frac{d\Gamma^{pow}}{dT} = -\frac{1}{E_{iz} + \gamma
    T_{target}}\frac{dP_{imp}}{dT_{target}} - \frac{P_{in} - P_{imp}}{\left(E_{iz} + \gamma
    T_{target}\right)^2} \left(\frac{dE_{iz}}{dT_{target}} + \gamma\right) .
  \label{eq:gammapowgradient}
\end{equation}
The gradient of $\Gamma^{pow}$ can therefore be changed by either
impurity power radiation which depends on target temperature, or through
an $E_{iz}$ which depends sufficiently strongly on target
temperature.

\begin{figure}[h]
  \centering
  \includegraphics[width=0.6\textwidth]{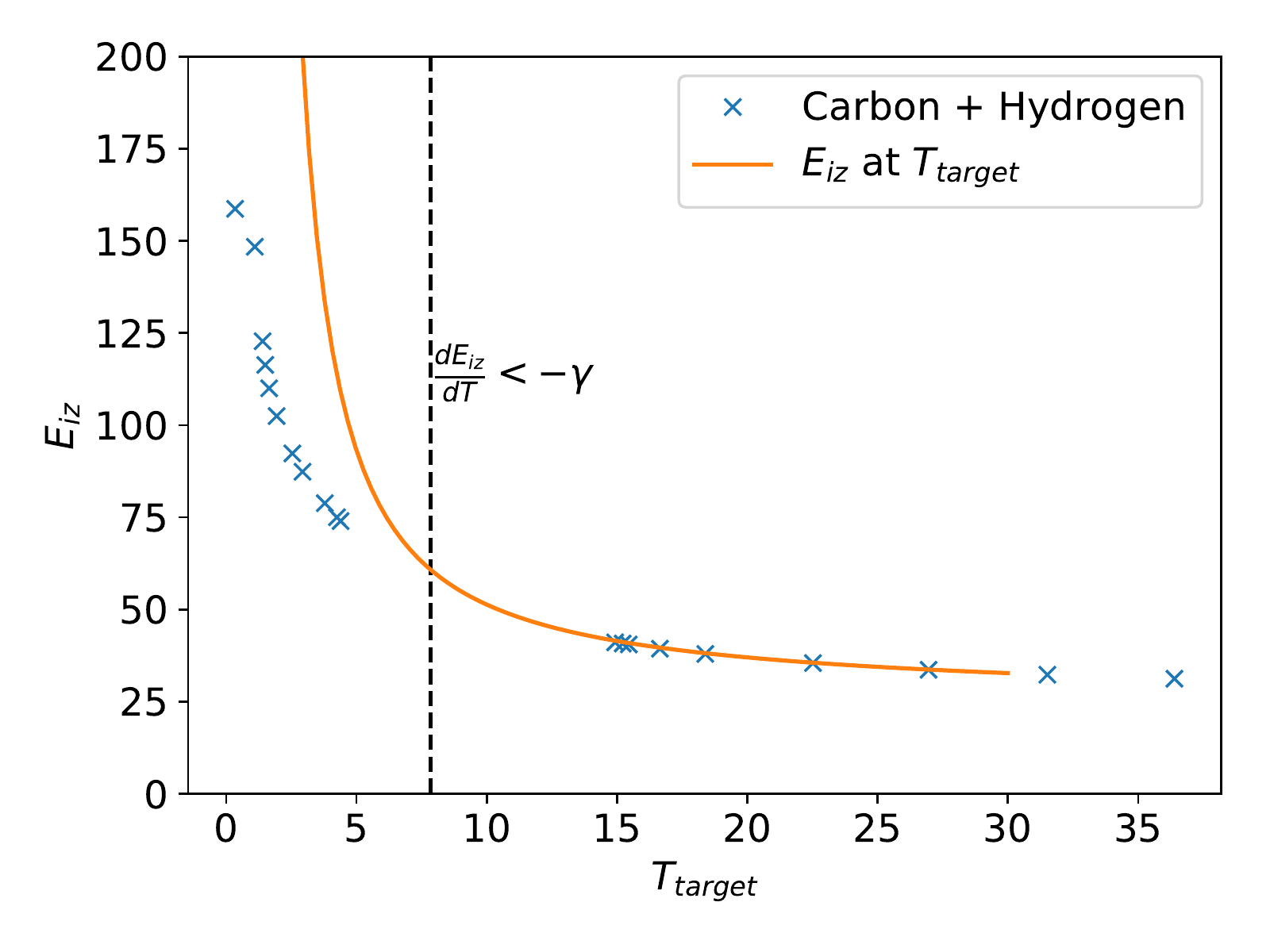}
  \caption{Effective $E_{iz}$ for the ``Carbon + Hydrogen'' case (Blue
    crosses),
    calculated as the ratio of the total power lost to hydrogen
    ionisation and excitation, divided by the rate of 
    ionisations, and averaged over the domain.
    The value of $E_{iz}$ from the rate coefficients used
    in SD1D at the target temperature $T_{target}$ are shown for
    comparison (red line). The threshold in equation~\ref{eq:eizthreshold} is
    marked by a vertical dashed line.}
  \label{fig:eiz_te}
\end{figure}

Impurity radiation can drive a change in sign of
$\frac{d\Gamma^{pow}}{dT}$, but is complicated by the dependence of
$P_{imp}$ on plasma density. Here we consider the case without
impurity radiation, $P_{imp}=0$. This is relevant to the scan labelled
``Carbon + Hydrogen'' since in those cases the hydrogen radiation
dominates the power loss. The criterion for a change in gradient of
$\Gamma^{pow}$, and so a rollover of target flux, is therefore given
by equation~\ref{eq:eizthreshold}:
\begin{equation}
  \left.\frac{\partial E_{iz}}{\partial T_{target}}\right|_{critical}  < -\gamma .
  \label{eq:eizthreshold}
\end{equation}
The effective value of $E_{iz}$ from the SD1D simulation with varying
$E_{iz}$ (labelled ``Carbon + Hydrogen'') is shown in figure~\ref{fig:eiz_te},
averaged over the field line for each case corresponding to the solution $T_{target}$. 
Since $E_{iz}$ is an increasingly steep function of temperature as the target
temperature falls, equation~\ref{eq:eizthreshold} can be related to a threshold
for rollover at $T_{target}=7.84$eV, $E_{iz}=60.8$eV in the red curve of
figure~\ref{fig:eiz_te} (point-wise value, not averaged over the domain). By
combining equations~\ref{eq:gammamom} and \ref{eq:powerbalance}, this threshold
in the gradient of $E_{iz}$ can also be related to a critical ratio of upstream
pressure to recycling power $p_{up} / P_{recl}$, which has been proposed as a
detachment threshold~\cite{krasheninnikov-1999,pschenov2017,verhaegh}, and which
is a function only of target temperature as shown in equation~\ref{eq:p_q}:
\numparts
\begin{eqnarray}
  p_{up}/P_{recl} &=&
  \frac{\sqrt{8m_iT_{target}}}{\left(1-f_{mom}\right)\left(E_{iz} + \gamma
                      T_{target}\right)} \label{eq:p_q}\\
                  &=& 12.6 \textrm{~N/MW} ,
\end{eqnarray}
\endnumparts where equation~\ref{eq:momlossfit} has been used for
$f_{mom}$. This value is smaller than the $\sim 17$N/MW threshold found
in~\cite{pschenov2017} for SOLPS4.3 simulations of pure Deuterium DIII-D-like
equilibria, but larger than that derived and measured~\cite{verhaegh}. This
difference may be due to different effective $E_{iz}$, which here uses a
simplified semi-analytic model, to differences in $\gamma$, or to a difference
in $f_{mom}$, which varies somewhat between devices and
simulations~\cite{stangeby2018}. The effective $E_{iz}$ averaged over the domain
(data points in figure~\ref{fig:eiz_te}) deviates from $E_{iz}$ at the target
temperature (curve in figure~\ref{fig:eiz_te}), which also shifts
equation~\ref{eq:p_q} to higher $p_{up}/P_{recl}$.

\begin{figure}[h]
  \centering
  \includegraphics[width=0.6\textwidth]{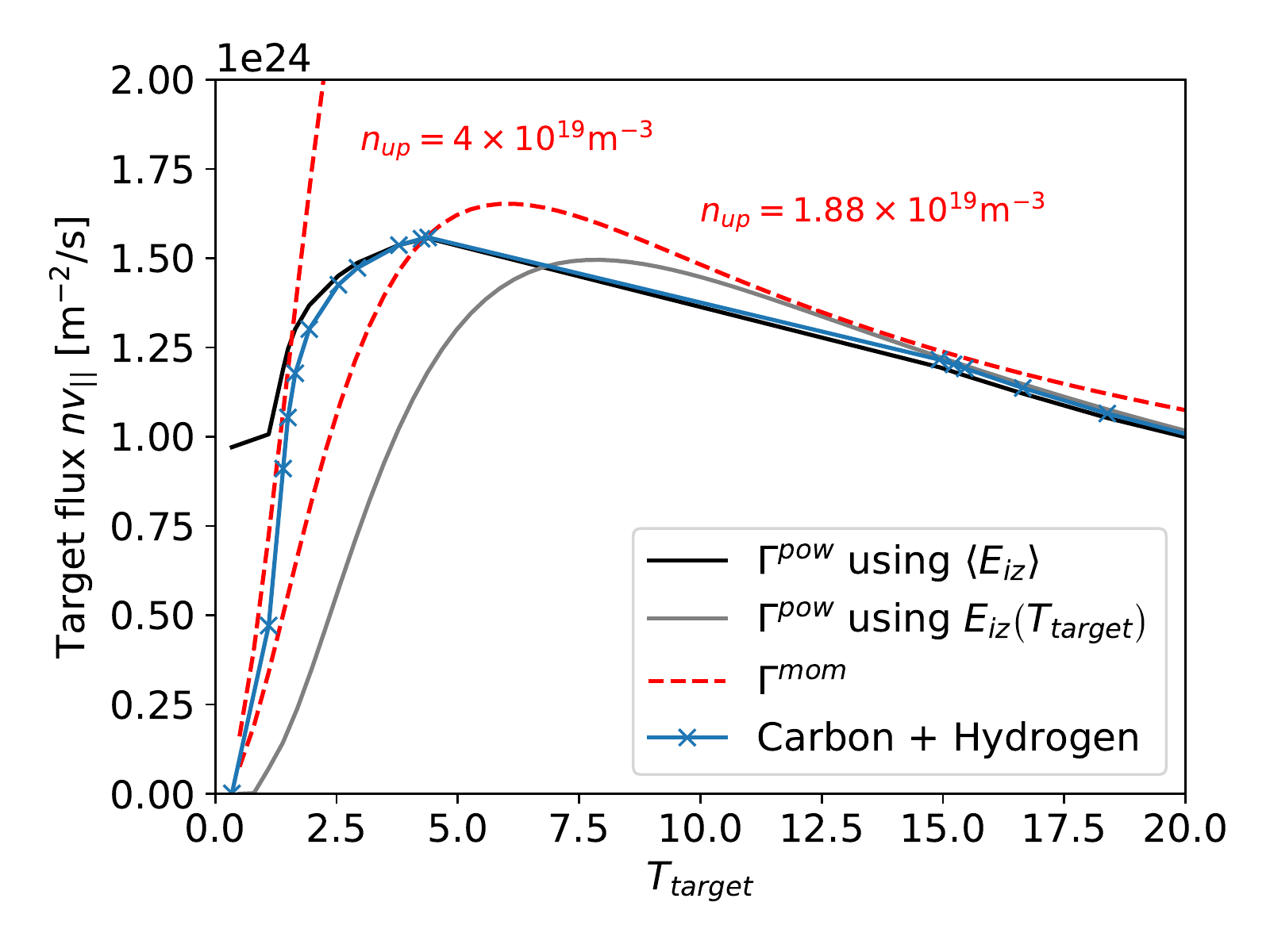}
  \caption{Target particle flux as a function of target temperature for the
    density scan with varying $E_{iz}$ (hydrogen excitation radiation) and 1\%
    carbon impurity. The grey solid curve shows $\Gamma^{pow}$ calculated using
    equation~\ref{eq:powerbalance} with fixed $P_{recl}$ and the value of
    $E_{iz}$ at $T_{target}$. The black solid curve shows $\Gamma^{pow}$
    calculated using the volume averaged $\left<E_{iz}\right>$ shown in
    figure~\ref{fig:eiz_te}.  Red dashed curves (again self-similar shape as
    $n_{up}$ is varied) show $\Gamma^{mom}$ calculated using
    equation~\ref{eq:gammamom}, with the fit for $f_{mom}$ from
    figure~\ref{fig:pratio} and equation~\ref{eq:momlossfit}.}
  \label{fig:teflux_eiz}
\end{figure}
The target flux for this scan with varying $E_{iz}$ (labelled ``Carbon
+ Hydrogen'' in figures~\ref{fig:area_2_assumptions} and~\ref{fig:norad}) is shown in
figure~\ref{fig:teflux_eiz}. This is similar to
figure~\ref{fig:teflux}, but now the black and grey curves which represent the
power balance constraint (equation~\ref{eq:powerbalance}) are not
monotonic and so a flux rollover is possible. The grey curve shows the
power balance obtained by using the target temperature in the $E_{iz}$
rates used in SD1D. This fits the simulation results well at high
target temperatures, but at an upstream density
$n_{up}\simeq 1.88\times 10^{19}$m$^{-3}$ a transition is seen in the target
temperature. Figure~\ref{fig:teflux_eiz} shows that the reason for
this transition is that at the corresponding target temperature $T_{target}=11.3$eV
the gradients of the grey and red curves are such that
$\frac{\partial\Gamma^{pow}}{\partial T} > \frac{\partial\Gamma^{mom}}{\partial T}$
so that these curves no longer intersect as the upstream density is
increased. This loss of steady-state solution results in a rapid change in
density and temperature profiles, approximately doubling the peak
density. This moves the peak in ionisation away from the target,
modifying the power dissipation curve (from the grey to the black line in
figure~\ref{fig:teflux_eiz}) with an average $E_{iz}$ which deviates
from the value at $T_{target}$, as shown in
figure~\ref{fig:eiz_te}. The solution moves to a different 
point on the same red curve at a similar upstream density
($1.88\times 10^{19}$m$^{-3}$ to $1.8825\times 10^{19}$m$^{-3}$),
but now with a target temperature of $3.8$eV rather than $11.3$eV.

Rapid transitions into detachment are common
(e.g.~\cite{lipschultz1996,krasheninnikov-1997,mclean-2015}) but are undesirable in a
tokamak divertor, both for design and control. The above discussion implies
that power dissipation mechanisms which dissipate strongly at high temperatures,
particularly at target temperatures above the peak in $\Gamma^{mom}$
(equation~\ref{eq:gammamom}), have the potential to result in sudden
transitions. This is because at these high temperatures
$\frac{\partial\Gamma^{mom}}{\partial T} < 0$ so that sufficiently strong power
dissipation can produce a situation where
$\frac{\partial\Gamma^{pow}}{\partial T} > \frac{\partial\Gamma^{mom}}{\partial
  T}$, the black and red curves in figure~\ref{fig:teflux_eiz} no longer
intersect, and a smooth change is no longer possible. For these simulations the
peak in $\Gamma^{mom}$ occurs at $6.0$eV when the fit in
equation~\ref{eq:momlossfit} is used. This analysis points to a possible
solution: mechanisms which remove momentum at higher target temperatures,
perhaps including radial transport, which modify
$f_{mom}\left(T_{target}\right)$ and increase
$\frac{\partial\Gamma^{mom}}{\partial T}$, would reduce the likelihood of sudden
transitions.

We conclude that when recombination is not significant,
momentum loss without sufficient power loss
results in a lack of target flux rollover (figure~\ref{fig:teflux});
power loss without sufficient pressure loss can
result in rapid transitions between states (figure~\ref{fig:teflux_eiz}).
More specifically, a change in gradient of $\Gamma^{pow}$ w.r.t
$T_{target}$ is needed for target flux rollover. If this change in
gradient is driven by hydrogen excitation radiation, then there is a
threshold in the gradient of $E_{iz}\left(T_{target}\right)$
(equation~\ref{eq:eizthreshold}) which can be related to a threshold
in $T_{target}$ (figure~\ref{fig:eiz_te}) and to a threshold in
$p_{up}/P_{recl} \simeq 12.6$~N/MW (equation~\ref{eq:p_q}). If this
threshold is reached at a higher temperature than the rollover in
$\Gamma^{mom}$ at $\sim 6$eV then a rapid transition may occur (figure~\ref{fig:teflux_eiz}), since a
smooth variation in plasma state is no longer possible.


\subsection{Removing charge exchange}

\begin{figure}[h]
  \centering
  \includegraphics[width=0.6\textwidth]{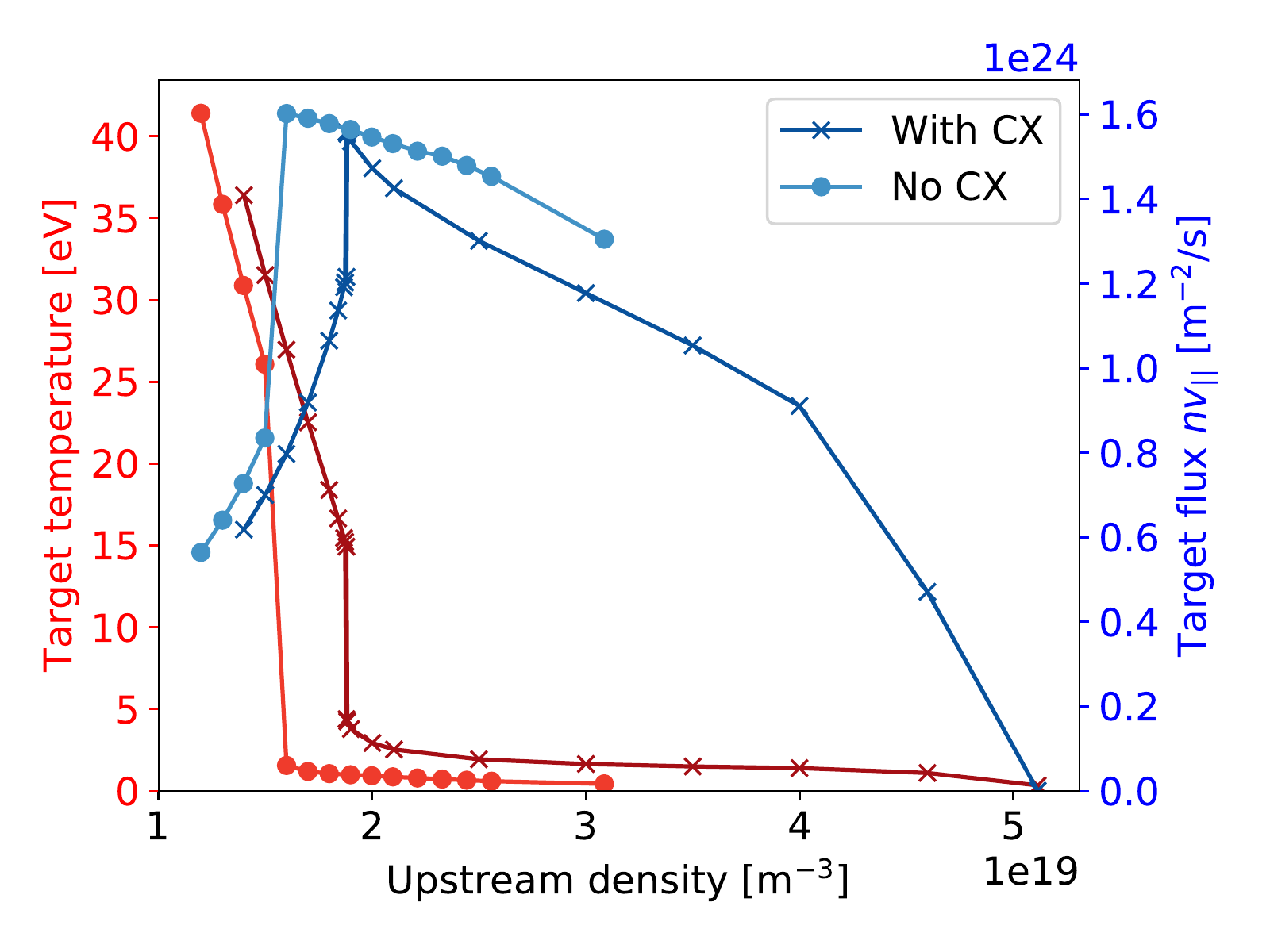}
  \caption{Effect of turning off charge exchange: Target temperature (red) and particle flux (blue) at a range of upstream densities. Area expansion factor $2$, input power $50$MW/m$^2$.}
  \label{fig:nocx}
\end{figure}

\begin{figure}[h]
  \centering
  \includegraphics[width=0.45\textwidth]{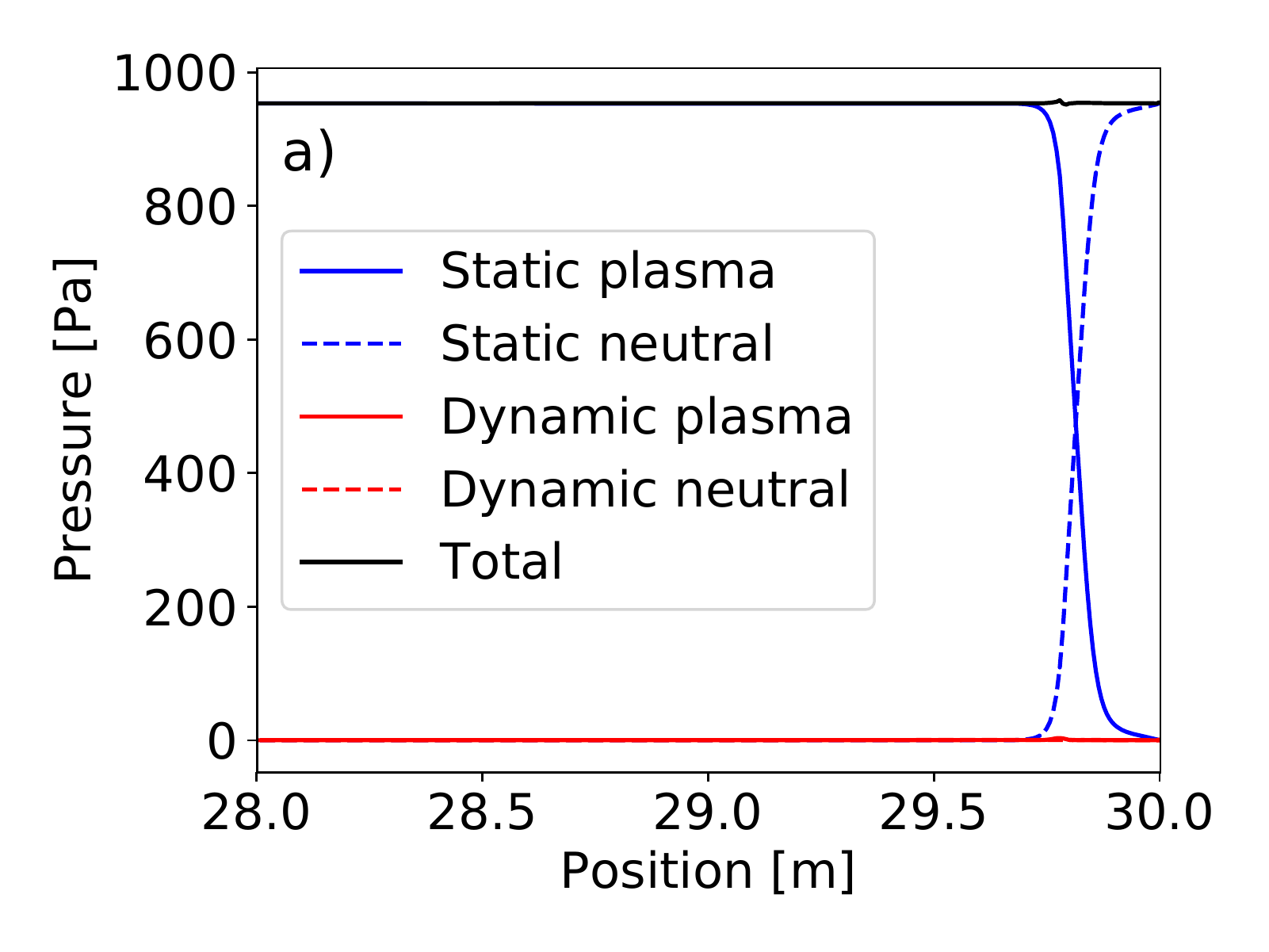}
  \includegraphics[width=0.45\textwidth]{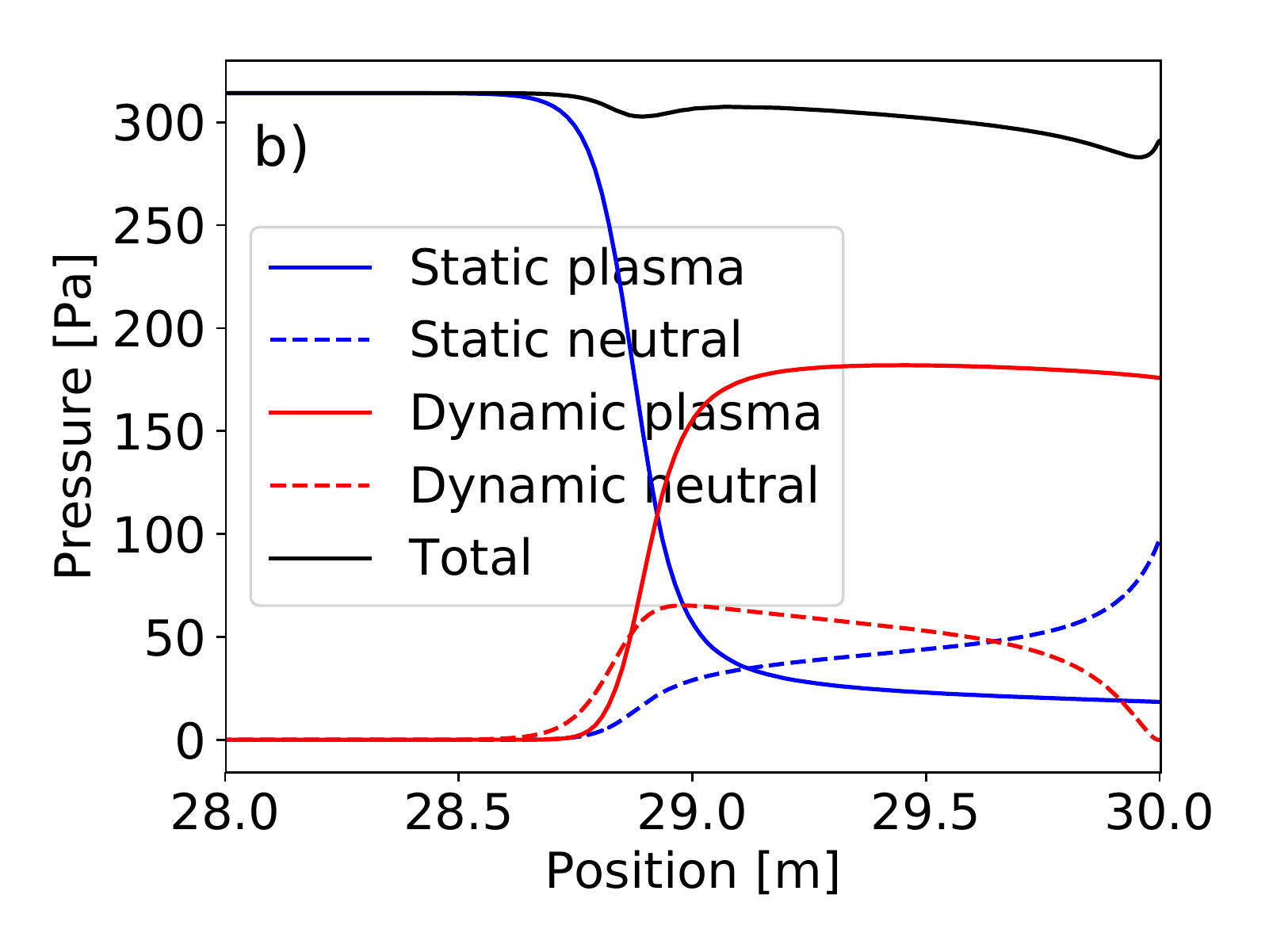}
  \caption{Parallel pressure balance a) with charge exchange at
    $n_{up}=5\times 10^{19}$m$^{-3}$ and b) without charge exchange at
  $n_{up}=1.7\times 10^{19}$m$^{-3}$.}
  \label{fig:pressure}
\end{figure}

In previous sections the effect of removing power dissipation
mechanisms has been studied, identifying the potential for transitions to
occur when insufficient momentum is removed. We now test the effect of
removing the dominant plasma momentum loss mechanism in these
simulations, charge exchange.
At rollover the pressure loss due to charge exchange is $16$
times larger than the sum of all other effects; the next most significant effect
is ionisation ($15.4$ times smaller) which acts in the opposite
direction. Given the discussion above of equation~\ref{eq:momentum2pt},
it might be expected that turning off charge exchange would raise
target flux and inhibit the roll-over during detachment. What is seen in
figure~\ref{fig:nocx} is more nuanced: At low densities the target flux is indeed
increased relative to the case with charge exchange, but the sharp
drop in target temperature and roll-over of target flux associated
with detachment is seen to occur at a lower upstream density.
The plasma profiles are qualitatively different in the cases with and
without charge exchange: When charge exchange is included,
conservation of momentum between 
plasma and neutral fluids results in a static solution shown in
figure~\ref{fig:pressure}a, in which the plasma fluid pressure is balanced by a neutral
cushion in front of the target: since neutrals do not escape, the total
plasma + neutral pressure is conserved. When charge exchange is turned
off, this static solution is not possible because the required momentum exchange cannot
occur. Instead in the solution shown in figure~\ref{fig:pressure}b a part of the plasma
momentum is lost to neutral pressure through ionisation ($102$~Pa) and
some through the magnetic mirror effect~\cite{ghendrih2011} ($20$~Pa), but the
majority of the upstream static plasma pressure ($313$~Pa) is balanced by
plasma dynamic pressure $m_inv_{||}^2$ ($172$~Pa). Power is radiated in a
quasi-neutral thermal front (gradient) region, and
acceleration of the plasma into the low pressure region
behind the thermal front converts internal energy to kinetic energy.
In the absence of momentum loss this results in a cold supersonic flow
of plasma to the target, reminiscent of a de Laval rocket nozzle~\cite{landau-lifshitz-fluids}.
Note that in this case anisotropic pressure (viscous) effects are
likely to be important~\cite{ghendrih2011}, but are not included here. The model
is quasi-neutral by construction, so if space charge effects play a role in the
thermal front then these would not be observed here.
The rarefaction front at which plasma static pressure drops moves
upstream as the upstream density is increased. This results in a
larger drop in the upstream temperature: Between upstream densities
$n_{up}=1.7\rightarrow 3\times 10^{19}$m$^{-3}$ the upstream temperature falls
from $58.8\rightarrow 58.2$eV if charge exchange is included, and
$58\rightarrow 43$eV if charge exchange is excluded. Over the same
range of upstream densities the front moves from $L_{||} \simeq 29$m
to $L_{||} \simeq 13$m, close to the X-point at $10$m.
The drop in upstream temperature is consistent with a simple 2-point
scaling~\cite{stangeby-2000} with 
parallel connection length $T^{upstream}\sim L_{||}^{2/7}$, where $L_{||}$ is
the location of the density peak. Such a use of the 2-point scaling treats the
density peak as a 'virtual target'~\cite{kukushkin-1997,stangeby-2011}.

The transition at lower upstream density in the case without charge
exchange, compared to the case with charge exchange, can be understood
by considering the power and momentum balance as done in
figures~\ref{fig:teflux} and \ref{fig:teflux_eiz}.
\begin{figure}[h]
  \centering
  \includegraphics[width=0.6\textwidth]{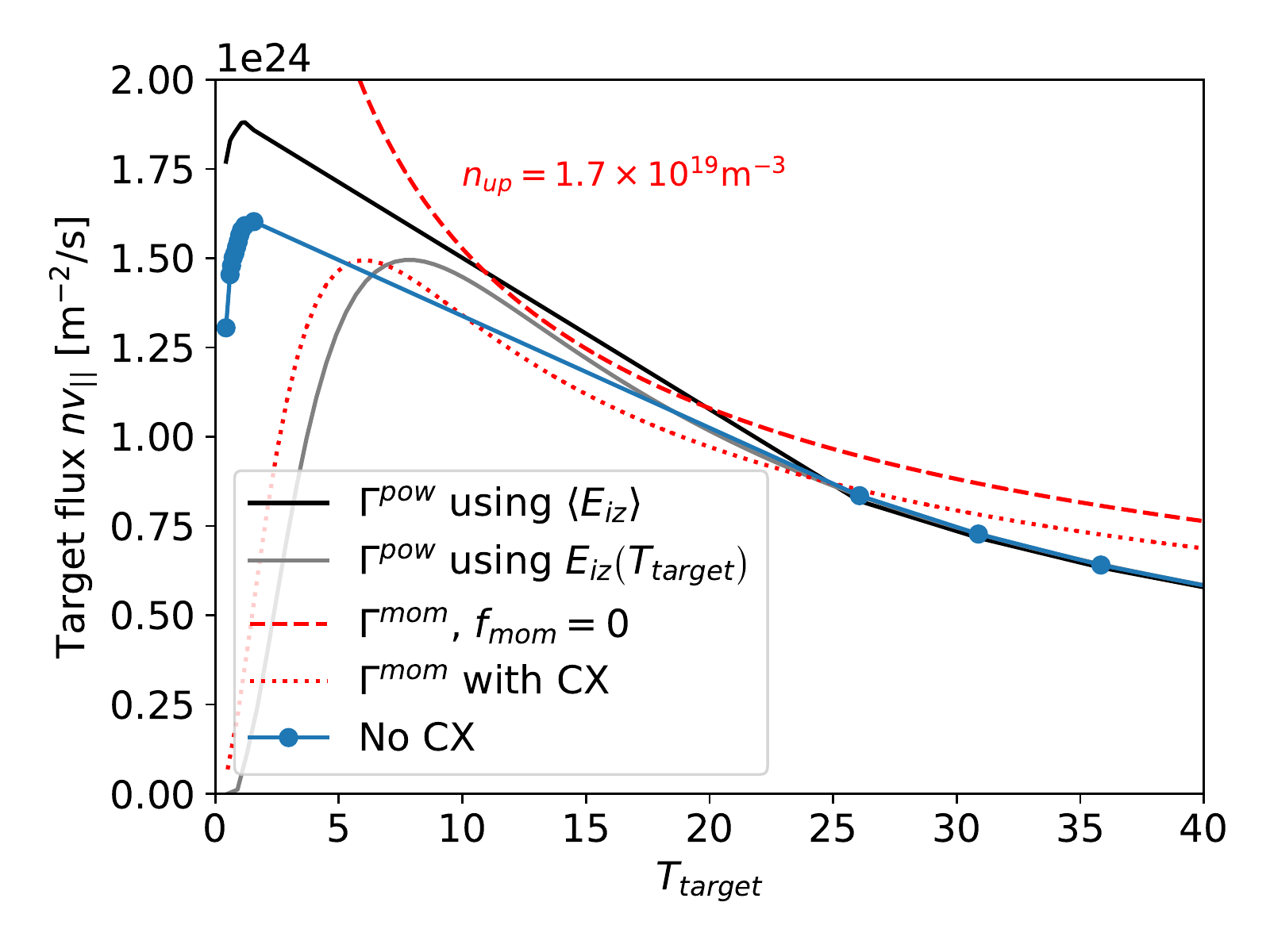}
  \caption{SD1D results for a case with no charge exchange (blue circles,
    labelled ``No CX''); Target particle flux calculated from power balance
    (eq~\ref{eq:powerbalance}) using $E_{iz}$ at the target temperature (grey
    solid line), and power balance using the average $E_{iz}$ over the
    simulation (black solid line). Dashed red lines show flux from momentum
    balance (eq~\ref{eq:gammamom}) assuming no momentum loss. At low upstream
    densities the momentum balance curve (dashed red) and power balance curve
    (grey solid) intersect the SD1D results. At upstream densities above
    $n_{up}\sim 1.7\times 10^{-3}$m$^{-3}$ the momentum balance curve does not
    intersect the power balance curve, and a transition is seen. For comparison
    the momentum balance curve for the case with charge exchange is shown as a
    dotted red line. At $n_{up}\sim 1.7\times 10^{-3}$m$^{-3}$ the dotted line
    does intersect the power balance curve and so no transition is seen until
    higher upstream density when charge exchange is included. }
  \label{fig:nocx_te_flux}
\end{figure}
 This is shown in
figure~\ref{fig:nocx_te_flux}: Removing charge exchange momentum
losses reduces $f_{mom}$ and so 
increases $\Gamma^{mom}$ (equation~\ref{eq:gammamom}). This moves the
location where the $\Gamma^{mom}$ curve intersects the $\Gamma^{pow}$
curve, so that the same target temperature occurs at a lower upstream
density. This lowers the upstream density at which a transition is
observed, relative to the case with charge exchange
(figure~\ref{fig:teflux_eiz}). Setting $f_{mom}=0$, the momentum
balance curves (red dashed lines) in figure~\ref{fig:nocx_te_flux}
cease to insersect the power balance curve at an upstream density of
around $n_{up}=1.7\times 10^{-3}$m$^{-3}$, at which point a transition
is observed in the SD1D simulation.

It seems unlikely that this solution with supersonic
flow could ever occur in tokamak experiments, though the possibility
of transitions to supersonic flow in divertor plasmas has been shown
previously~\cite{ghendrih2011}. The dramatic change in
plasma solutions when momentum exchange is modified demonstrates the importance
of momentum exchange to determining the divertor detachment thresholds
and dynamics. It also shows the utility of the analysis in
figure~\ref{fig:nocx_te_flux} to explaining the observed simulation results.

\section{Conclusions}

We have undertaken steady-state, 1D, simulations of divertor plasma detachment, specifically target ion flux rollover, utilizing a new computational tool, SD1D. The importance of recombination, radiative power loss, and momentum exchange in the detachment process with emphasis on the behavior of the target ion current rollover has been evaluated.  

We find that for MAST-Upgrade like simulation parameters
recombination does not play a significant role at flux rollover
(target temperature $T_{target}\sim 3-5$eV), though is significant when
$T_{target}$ drops to $\sim 1$eV.

It is shown that in these simulations momentum loss, as characterised by a drop
in total pressure along the magnetic field, is insufficient, by itself, to
produce flux rollover during detachment. Impurity and/or hydrogenic radiative
losses are also required to increase at low temperatures. In the
particular case studied where excitation power losses dominate over impurity
radiation, this corresponds to $dE_{iz}/dT_{target} < -\gamma$. We conclude that when
recombination is not significant, momentum loss without sufficient power loss
results in a lack of target flux rollover.

The precise dependence of momentum and power losses on target temperature has
implications for the availability of steady state solutions. We have found that
when the target flux is characterized in terms of power balance,
$\Gamma^{pow}\lr{T_{target}}$, and momentum balance,
$\Gamma^{mom}\lr{p_{up}, T_{target}}$, the allowed intersections of the two
curves readily predict available steady state solutions. In some cases
(e.g. power dissipation occurring without sufficient momentum removal) we find
that there are regions of $T_{target}$ where there are no solutions – leading to rapid
drops in temperature for essentially no increase in upstream density.

One implication of this work is that modification of the shape of
$\Gamma^{pow}\lr{T_{target}}$ and $\Gamma^{mom}\lr{p_{up}, T_{target}}$ can be
used to influence detachment behavior. One can envision causing such shape
modifications by changing the escape probability of CX neutrals carrying
momentum (e.g. changing divertor target geometry) or changing the temperature
dependence of radiative losses (e.g. switching from one seeded impurity to
another).

The SD1D code is available at \texttt{https://github.com/boutproject/SD1D}.
All inputs, data and processing scripts are available at DOI \texttt{10.5281/zenodo.1410281}.

\section*{Acknowledgements}

This work has received funding from the EPSRC under grant
EP/N023846/1.  B. Lipschultz was funded in part by the Wolfson
Foundation and UK Royal Society through a Royal Society Wolfson
Research Merit Award as well as by the RCUK Energy Programme (EPSRC
grant number EP/I501045).

This work has been carried out within the framework of
the EUROfusion Consortium and has received funding
from the Euratom research and training programme
2014-2018 under grant agreement No 633053. The views
and opinions expressed herein do not necessarily reflect
those of the European Commission

\section*{References}
\bibliography{references}
\bibliographystyle{unsrt}

\end{document}